# Modularity of the metabolic gene network as a prognostic biomarker for hepatocellular carcinoma

Fengdan Ye[1,2]*, Dongya Jia[2,3]*, Mingyang Lu[4], Herbert Levine[1,2,5,6,#], Michael W Deem[1,2,3,5,#]

[1]Department of Physics & Astronomy, Rice University, Houston, TX 77005, USA
[2]Center for Theoretical Biological Physics, Rice University, Houston, TX 77005, USA
[3]Program in Systems, Synthetic and Physical Biology, Rice University, Houston, TX 77005, USA
[4]The Jackson Laboratory, Bar Harbor, ME 04609, USA
[5]Department of Bioengineering, Rice University, Houston, TX 77005, USA
[6]Department of Biosciences, Rice University, Houston, TX 77005, USA
*co-first authorship
#corresponding author

**Corresponding author: Dr. Michael W Deem**

Department of Bioengineering, Rice University, Houston, TX 77005, USA

Phone: 713-348-5852 Fax: 713-348-5811

Email: mwdeem@rice.edu

**Corresponding author: Dr. Herbert Levine**

Department of Bioengineering, Rice University, Houston, TX 77005, USA

Phone: 713-348-8122 Fax: 713-348-8125

Email: herbert.levine@rice.edu

**The authors declare no conflicts of interest.**

**Abstract**

Abnormal metabolism is an emerging hallmark of cancer. Cancer cells utilize both aerobic glycolysis and oxidative phosphorylation (OXPHOS) for energy production and biomass synthesis. Understanding the metabolic reprogramming in cancer can help design therapies to target metabolism and thereby to improve prognosis. We have previously argued that more malignant tumors are usually characterized by a more modular expression pattern of cancer-associated genes. In this work, we analyzed the expression patterns of metabolism genes in terms of modularity for 371 hepatocellular carcinoma (HCC) samples from the Cancer Genome Atlas (TCGA). We found that higher modularity significantly correlated with glycolytic phenotype, later tumor stages, higher metastatic potential, and cancer recurrence, all of which contributed to poorer overall prognosis. Among patients that recurred, we found the correlation of greater modularity with worse prognosis during early to mid-progression. Furthermore, we developed metrics to calculate individual modularity, which was shown to be predictive of cancer recurrence and patients' survival and therefore may serve as a prognostic biomarker. Our overall conclusion is that more aggressive HCC tumors, as judged by decreased host survival probability, had more modular expression patterns of metabolic genes. These results may be used to identify cancer driver genes and for drug design.

**Introduction**

Hepatocellular carcinoma (HCC) is a primary malignancy of the liver, with average survival time between 6 to 20 months without any intervention [1]. It is also the third leading cause of cancer mortality worldwide [2]. The prognosis for HCC patients remains poor [3]. Diagnosis of HCC is usually based on biomarkers, such as AFP (alpha-fetoprotein) and miR-21 [4]. However, HCC can result from a variety of risk factors, such as hepatitis B/C virus or alcoholic liver disease [5], which makes it difficult to characterize HCC with single gene biomarkers. One key to a further breakthrough in HCC therapy lies in better understanding the underlying mechanism of HCC progression.

In recent years, a significant amount of research has gone into analyzing cancer-associated pathways and networks to gain insight into the complex biological systems underlying tumor progression [6, 7]. One promossing approach for breast cancer and leukemia patients has been to identify the varying patterns of cancer-associated gene expression to predict prognosis [8, 9]. In both of these examples, the level of organization of the cancer-associated gene network, as measured by the cophenetic correlation coefficient, (CCC), was shown to be correlated with cancer risk, progression, and outcome. Inspired by these works, we here aim to characterize HCC progression and patient survival by analyzing the structure of the HCC cancer-associated gene network.

Liver is an organ in which metabolism plays a key role. And abnormal metabolism is a hallmark of cancer [10, 11]. Therefore, we chose to analyze the structure of metabolic gene expression of HCC patients. Unlike normal cells, cancer cells use glycolysis for energy production irrespective of the availability of oxygen, a process termed the Warburg effect or aerobic glycolysis [12, 13]. Interestingly, although aerobic glycolysis has been regarded as the dominant metabolism phenotype in cancer, recent experimental evidence shows that mitochondria are actively functional in cancer cells [14-16], and oxidative phosphorylation (OXPHOS) can enhance metastasis in certain scenarios [17, 18]. Study of the interplay between glycolysis and OXPHOS will deepen our understanding of cancer metabolism and metastasis.

To quantify the activities of the two main metabolism phenotypes in HCC, OXPHOS and glycolysis, Yu *et al.* [19] developed AMPK and HIF-1 signatures by evaluating the expression of the downstream genes of AMPK (5' AMP-activated protein kinase) and HIF-1 (hypoxia-inducible factor 1), in total 33 AMPK downstream genes and 23 HIF-1 downstream genes. The AMPK and HIF-1 signatures have been shown to capture metabolic features of HCC samples [19]. In addition, the AMPK and HIF-1 signatures can associate the metabolism phenotypes of HCC samples with oncogene activities, such as MYC, c-SRC and RAS, which further validates the use of the AMPK and HIF-1 signatures in characterizing the metabolic activity of HCC samples [19]. Based on these arguments, the AMPK and HIF-1 downstream genes were chosen for the present study as a relevant set of cancer-associated genes. A strong anti-correlation between AMPK and HIF-1 activity was observed in HCC, suggesting the expression of these metabolic genes is modular, with the AMPK and HIF-1 downstream gene subsets as two likely modules (see Results section).

Community structure of a gene network conveys information regarding the interaction between genes. In particular, genes within the same community cooperate much more with each other than with those in other communities. By investigating the community structure of the HCC cancer-associated gene network, we identified key changes predictive for tumor progression and patient death. We found that higher community structure significantly correlated with glycolytic phenotype, later tumor stages, higher metastatic potential, and cancer recurrence, all of which contributed to poorer overall prognosis. Among patients that recurred, we found the correlation of greater community structure with worse prognosis during early to mid-progression.

Here we utilize modularity to quantify the community structure of the HCC cancer-associated gene network. Modularity is a measure of intracommunity connection strength compared to what is expected from randomly distributed connections [20, 21]. In the current context, it quantifies the ability of tumor cells to organize individual cancer-associated genes so as to maximize network efficiency. Modularity is present in almost all biological systems, from molecular interactions to macroscopic food webs [22, 23]. A general theory regarding modularity shows that high modularity systems afford greater evolutionary fitness in high stress environments or over shorter time scales, whereas low modularity systems afford greater fitness in low stress environments or over longer time scales [24, 25]. This general principle can be applied to understand the relation between modularity of cancer-related gene networks and the aggressiveness of cancer [8, 9]. Using this theory, we predict that tumors with a more modular expression pattern of cancer-associated genes, organized to counteract host defenses, are more fit and aggressive. At longer time scales, tumor growth overcomes host defenses and loses its sensitivity to host actions, and modularity is predicted to decline.

In this work, we analyzed the change of the modular expression pattern of the AMPK and HIF-1 downstream genes in HCC samples as a function of metabolism phenotypes, tumor stages, metastatic potentials, and recurrence status. We found that (i) HCC samples with a glycolysis phenotype show significantly higher modularity than samples with an OXPHOS phenotype; (ii) HCC samples at tumor stages II-IV have significantly higher modularity than samples at stage I; (iii) HCC samples with higher metastatic potential maintain significantly higher modularity than samples with lower metastatic potential; and (iv) patients that have recurrence within 12, 24 or 36 months have significantly higher modularity than those with no recurrence within the same amount of time. These results confirm the theoretical prediction that more aggressive tumors correspond to a more modular interaction pattern of the cancer-associated gene network. We also found that modularity increases with tumor progression up to 8 months before recurrence, but then decreases. This result is examined in detail in the Discussion section, and indicates that modularity is no longer selected for at very late stages of tumor progression. This result is also in accord with the aforementioned theoretical expectations. We further developed metrics to calculate individual modularity, which proved to be predictive of recurrence and survival for individual HCC patients. Possible applications of modularity in terms of drug design and identifying cancer-related genes will be discussed in the Discussion section.

**Results**

To construct our HCC cancer-associated gene network, we took the 33 AMPK downstream genes and 23 HIF-1 downstream genes identified by Yu et al. (2017) [19] as nodes in the network. For each group of patients, the interaction patterns between genes were calculated using Pearson correlation. Simply put, two genes have a strong interaction if they show a similar trend of gene expression changing across patients. That is, if one gene expression increases and another gene expression also increases, then these genes are cooperating and strongly interacting with each other. After nodes and links were established, we applied the Newman algorithm [21] to obtain the community structure of the gene network and the corresponding modularity value.

**Modular expression pattern of the AMPK and HIF-1 downstream genes**.

There exists a strong anti-correlation between the AMPK activity and HIF-1 activity across all 371 samples (Fig. 1A). In addition, expression of individual AMPK downstream genes was highly positively correlated within the AMPK gene group and negatively correlated with the HIF-1 downstream genes, and *vice versa* (Fig. 1B). The expression pattern of these genes was highly modular and consisted of two modules, one containing mainly AMPK-downstream genes and the other HIF-1-downstream genes, as identified by the Newman algorithm (Fig. 1C).

**Modularity and metabolism phenotypes.**

To evaluate the modular gene expression pattern of different metabolism phenotypes of HCC samples, we performed principal component analysis (PCA) on the RNA-Seq data of 33 AMPK downstream genes and 23 HIF downstream genes. Since AMPK and HIF-1 are master regulators of OXPHOS and glycolysis, respectively [19], the resulting first principal components (PC1s) for AMPK and HIF-1 downstream genes were assigned as the axes to quantify the activities of OXPHOS and glycolysis. After projecting all 371 HCC samples to the AMPK and HIF-1 axes, each cancer sample was assigned a metabolic state of glycolysis (HIF-1$^{high}$/AMPK$^{low}$), hybrid (HIF-1$^{high}$/AMPK$^{high}$) or OXPHOS (HIF-1$^{low}$/AMPK$^{high}$) through $k$-means clustering using the sum of absolute differences (Fig. 2A). Group modularity calculation showed that the OXPHOS group had the lowest mean modularity, the hybrid group had an intermediate mean value of modularity, and glycolysis group had the highest mean modularity (Fig. 2B). Combined with survival curves of the three groups (Fig. 2C), it is clear that the glycolysis group had the worst survival and OXPHOS the best, with hybrid in the middle, indicating that higher modularity corresponded to a more aggressive tumor.

**HCC samples at later tumor stage have greater modularity**

To analyze the change of modularity with respect to tumor stage, we classified the 348 of the 371 HCC samples that have neoplasm disease stage information into two groups, stage I (171 samples) and stage II-IV (177 samples). This was done to ensure that each group has similar number of samples. Group modularity calculations show that the stage II-IV group had a significantly higher mean modularity than the stage I group (Fig. 3A). Stage II-IV samples also had a significantly worse survival than stage I samples (Fig. 3B), which further confirmed that higher modularity corresponded to worse survival, i.e. a more aggressive tumor.

**HCC samples with higher metastatic potential have greater modularity**

Metastasis accounts for more than 90% cancer related deaths [26]. To evaluate the correspondence of modularity to metastatic potential of HCC samples, we grouped the samples based on their metastatic potential and calculated the group modularity. Genes SNRPF, EIF4EL3, HNRPAB, DHPS, PTTG1, COL1A1, COL1A2, LMNB1 (comprising the eight-gene signature) have been shown to be upregulated in metastases compared to primary tumor sites [27]. Expression levels of these genes has been used to evaluate the metastatic potential of primary tumors [27]. We here used the sum of log2-transformed values of expression levels of these eight genes to represent the metastatic potential of primary HCC samples. The 123 samples with the lowest metastatic potential were classified as the low potential group, and the 123 samples with the highest metastatic potential as the high potential group. Group modularity calculation results show that the high metastatic potential group had higher modularity and worse prognosis (Fig. 4A). We also used the expression of gene SPP1 to quantify the metastatic potential of HCC samples since the single SPP1 gene has been shown to be a diagnostic marker for metastatic HCC [28]. The grouping of HCC samples by expression of SPP1 show consistent results to that observed from the eight-gene signature (Fig. 4B). This result indicated that a highly modular pattern of cancer-associated gene interactions is a sign of tumor metastasis.

**Modularity and tumor recurrence**.

Tumor relapse is a supreme clinical challenge [29]. To analyze how prognosis of tumor relapse in connected to the modularity of metabolic genes in HCC samples, we classified the 319 of 371 HCC samples that have tumor recurrence information – 'recurred' or 'disease free'. Here the 319 samples were classified into non-recurrence and recurrence groups within 12 months, 24 months, or 36 months. For example, the recurrence group within 12 months includes HCC samples whose disease-free status was 'recurred' and the 'disease free time' was shorter than 12 months. The non-recurrence group within 12 months includes HCC samples whose 'disease free time' was longer than 12 months, with either 'recurred' or 'disease free' status.

In all three cases, we observed that the group of patients that recurred had a higher mean modularity than the group of patients that did not (Fig. 5A). The difference between the recurrence and no-recurrence groups became more significant as time increased from 12 to 24 to 36 months. The survival curves confirmed that the recurrence group, which was also the high modularity group, had poor survival (Fig. 5A).

To understand the origin of the correlation between higher modularity and worse survival, we examined the relation between modularity and tumor recurrence time among recurred patients. Among the 319 samples, 174 have disease-free status as 'recurred'. After discarding the 4 patients with the longest disease free time, the rest were sorted based on the disease-free time and classified into 5 groups – group 1, 2, 3, 4 and 5 with decreasing disease-free time. That is, group 1 was the longest from recurrence, and group 5 was the nearest. The result is shown in Fig. 5B and the corresponding survival curves for each group are shown in Fig. 5C. Modularity first increased with tumor progression, and then decreased. The differences between each group, however, were not always significant. It is also worth noting that modularity correlated with worse survival for the first 3 groups, but the correlation is reversed for groups 4 and 5. This result is similar to the trend observed in a study of acute myeloid leukemia [9]. In early stages, increased modularity correlates with decreased survival time as cancer cells organize their gene expression against the host. In later stages, cancer has overcome the host defenses, and a high value of modularity is no longer selected for. Host survival, while low, becomes independent of modularity. We note that this crossover occurs rather late: recurrence times for groups 1, 2, 3 were 90-22 months, 22-13 months, 13-8 months; the recurrence times for groups 4 and 5 were 8-4 and 4-1 months, respectively. Note that Fig. 5B and 5C are based on recurred patients only, whereas Fig. 5A contains both recurred patients and disease-free patients.

**Clinical application of modularity: Individual modularity and prediction.**

Calculation of group modularity is useful for understanding the group differences of metabolic gene expression patterns and the general relation between modularity and malignancy. However, for clinical application, individual modularity is required in order to make predictions regarding individual prognosis. The detailed definition and calculation procedure of individual modularity can be found in the 'Materials and Methods'. Simply put, we applied the Newman algorithm to an individual cancer-associated gene network, with a new method to define links and with an additional de-noising step.

Individual modularity for all 371 samples ranged from 0.248 to 0.652, with mean 0.453 and standard deviation 0.079. These numbers appeared to be consistent with modularity found in other functional biological networks in human [30]. Modularity at the individual level largely confirmed the above group level trends of classification into OXPHOS or glycolytic metabolism, stage, recurrence status and metastatic potential. Higher individual modularity corresponded to the glycolysis phenotype, later tumor stage, and higher metastatic potential, as determined by the aforementioned eight-gene signature (Fig. 6A-C). Results regarding recurrence and SPP1 metastatic potential can be found in Supplementary Fig. S2. In all three cases, the group of patients that recurred always had a higher mean modularity than the group of patients that did not (Supplementary Fig. S2A-C). Similar to the case of the eight-gene signature of metastatic potential, a positive Pearson correlation was observed between individual modularity and metastatic potential calculated from the log2 transformation of the SPP1 expression (r=0.35, $p<10^{-11}$, Supplementary Fig. S2D). Together, these results validate the use of the metric of individual modularity to evaluate the aggressiveness of individual HCC patients.

To make prognostic predictions with individual modularity, we focus on survival and recurrence. We attempted to predict the probability of survival longer than 24 months and no recurrence in 12 months, so that each group has a comparable amount of samples (survived longer than 24 months: 140 samples; shorter than 24 months: 91 samples; no recurrence in 12 months: 176 samples; recurrence within 12 months: 104 samples). We then divided patients into 6 groups based on their individual modularity values (0.24-0.31, 0.31-0.38, 0.38-0.45, 0.45-0.52, 0.52-0.59). For each group, we counted the number of patients that survived longer 24 months and that remained disease-free for more than 12 months. We then calculated the proportion of these patients in each group (Fig. 6D and 6E, left panel). Overall, the higher the modularity, the lower the survival and disease-free probability. The only exception is the first bar in Fig. 6D left panel, which could potentially due to the very small number of patients in the group (7 patients).

We then captured these results with a Gaussian model of the modularity distribution of each group, with mean and standard deviation computed from individual modularity values of each group, Fig. 6D and 6E, middle panel. Based on (eq.4) and (eq.5) defined in the 'Materials and Methods' section, the probability of survival over 24 months and the probability of no recurrence in 12 months was calculated, Fig. 6D and 6E, right panel. This simple model was able to recapitulate the trend observed in the real data, Fig. 6D and 6E, left panel. The modularity range in these two plots was selected as 0.248 – 0.652 to match with the sample individual modularity values.

Individual modularity showed significant potential as a predictor of patient survival or recurrence. As was the case for the group modularity observations, a high value of individual modularity was suggestive of poor prognosis, with values of $M > 0.6$ correlated to survival and non-recurrence probabilities less than 0.4.

**Discussion**

Metabolic reprogramming is an emerging hallmark of cancer [10, 11]. Both aerobic glycolysis and oxidative phosphorylation (OXPHOS) play important roles in orchestrating cancer metabolism [12-15, 17, 18, 31]. Previously, Yu et al. developed the AMPK and HIF-1 signatures to quantify the activities of metabolism phenotypes in hepatocellular carcinoma (HCC) [19]. There was a visually apparentmodular pattern of gene expression due to the strong anti-correlation between AMPK and HIF-1 activity in HCC. In this work, we analyzed the gene expression pattern of metabolic genes in HCC in term of modularity and studied its correlation with metabolism phenotypes, tumor stages, metastatic potentials, and tumor recurrence.

The analyses of modularity in different metabolism phenotypes of glycolysis, hybrid, or OXPHOS; different stages of stage I or stage II-IV; varying tumor metastatic potentials; and differing recurrence statuses consistently showed that a higher modularity of the AMPK and HIF-1 downstream gene network corresponded to worse overall survival results for HCC patients. For example, a group of samples characterized by high glycolytic activity showed significantly higher modularity than a group of samples characterized by high OXPHOS activity, and worse prognosis. The result is consistent with the experimental observation that hepatocarcinogenesis initiates with a switch of metabolism from OXPHOS to glycolysis, and glycolysis is maintained to facilitate the aggressive features of advanced HCCs [32, 33]. Similarly, comparison of stage I patients to stage II-IV patients showed that the latter has a more modular expression of metabolic genes and worse survival prognosis. Additionally, patients with a higher metastatic potential had a more modular expression of metabolic genes and worse survival prognosis. Finally, those patients that recurred within a given time had a more modular expression of metabolic genes and worse survival prognosis that those that did not.

One interesting phenomena is a non-monotonic relation between modularity and tumor progression as shown in Fig. 5B. Modularity increased first and then decreased. This result is similar to the trend observed in a study of acute myeloid leukemia [9]. We argue that at early stages of tumor progression, a modular pattern of cancer-associated gene interactions is organized by tumor cells, so that they can counteract the host defense systems. At later stages of tumor progression, cancer has overcome the host defenses, and a high value of modularity is no longer selected for. The results here suggest that the relation between modularity and tumor aggressiveness is mediated by tumor progression. For most of the patientl's history, a higher modularity indicates higher risk. Only when tumor progression has reached a very late stage, may a lower modularity indicate higher risk. Therefore, an accurate interpretation of modularity should take progression stage into consideration.

We further investigated the relation between modularity and recurrence time of recurred patients in three patient subsets: glycolysis phenotype, stage II-IV, and high eight-gene metastatic potential, see Supplementary Fig. S3. These groups were chosen as demarcations of stress. The glycolysis group had 37 patients that recurred. After discarding 2 samples with the shortest recurrence time, the rest were distributed into 5 equal size groups. A similar procedure was taken for all groups in which the number of recurred patients was not divisible by 5. Again, group 1 was the longest from recurrence, and group 5 was the nearest. Survival curves for each group were also plotted. We find the correlation of greater modularity with worse prognosis for survival exists for the roughly ~60% (top 3 groups) of patients with the longest recurrence time in all cases. Interestingly, a reversal of this correlation occurs at around the same time for the different measures of stress: 9.1 months for the glycolysis group, 6.4 months for the stage II-IV group, and 7.9 months for the high metastatic potential group. These times are consistent with the reversal of the correlation at 8 months found among all recurred patients.

Taken together, these results show that modularity is selected for under the stressful conditions of early to mid-progression. That is, more aggressive tumors in these early and mid-conditions, as judged by decreased host survival probability, have greater modularity of metabolic genes. These results confirm our previous hypothesis that more malignant tumors are usually characterized by a more modular expression pattern of cancer-associated genes [8, 9]. We predict that higher modularity increases the fitness of tumors because metabolic networks are typically under increased stress in HCC tumor cells [34]. Thus, tumors with a more modular metabolic gene network typically are more fit, and it is these tumors that are able to overcome the body's defenses. Once the transition to imminent recurrence is achieved, the selection strength for modularity is no longer present, and the observed values of modularity decrease.

To the best of our knowledge, this is the first effort to evaluate aggressiveness of HCC samples by evaluating expression patterns of metabolic genes in terms of modularity. Further work can extend the modularity concept to different types of tumors. At least two avenues for improvement of the present results may be possible. First, a different set of parameters used in calculating individual modularity might affect the predictive efficiency. We list in Supplementary Table S2 the parameters for calculating individual modularity using ITSPCA. Variation of these standard parameter values gave similar results, but with a weaker signal. We therefore believe that the chosen parameter set works well and keeps most of the signal. Future work could look into further quantifying the signal as a function of the parameter set to improve the predictive power of individual modularity. Second, the temporal expression profiles of the metabolic genes in HCC samples from individual patients may further power the personalized predictions for outcome.

In summary, modular interactions between metabolic genes in HCC play a key role in HCC prognosis. Individual HCC patients with higher modularity have a higher risk of death and recurrence. These results for modularity may translate to a number of possible clinical applications: (i) prediction of patient survival and recurrence probabilities with individual modularity can enable the choice of appropriate therapies; (ii) key genes promoting HCC progression could be potentially identified, e.g. a hub node gene that strengthens intracommunity interactions and increases modularity is presumably an important cancer driver gene that helps facilitate tumor progression; and (iii) drug treatment efficacy could be evaluated by testing the ability of drugs to disrupt the modular interactions between cancer-associated genes. A novel approach to drug design could target genes that significantly contribute to the modularity of the cancer-associated gene network.

**Materials and Methods**

**1. 371 primary HCC samples**

RNA-Seq data for 373 hepatocellular carcinoma (HCC) samples, which contain the mRNA expression of 33 AMPK downstream genes and 23 HIF-1 downstream genes, were obtained from TCGA at cBioPortal. Among the 373 HCC samples, 371 primary tumor samples were used for subsequent analysis, and 2 recurrent tumor samples were excluded. More details can be found in Supplementary Materials and Methods section 1.

**2. Calculation of Group Modularity**

Modularity of a given graph $A_{ij}$ was defined as

$$M_G = \frac{1}{2e} \sum_{\substack{all\ module \\ areas\ i,j}} \sum_{\substack{within \\ this\ module}} \left(A_{ij} - \frac{a_i a_j}{2e}\right) \qquad (eq.1)$$

where $A_{ij}$ is 1 if there is an edge between nodes $i$ and $j$ and 0 otherwise, the value of $a_i = \Sigma_j A_{ij}$ is the degree of node $i$, and $e = \frac{1}{2} \Sigma_i a_i$ is the total number of edges. This definition can be extended to unsigned weighted graphs, where $A_{ij}$ is the weight of the edge between nodes $i$ and $j$ and where $A_{ij} > 0$. Here the subscript 'G' is used because this definition is adopted for calculation of group modularity. We applied Newman's algorithm [21] to graph $A_{ij}$ to calculate modularity. This algorithm found the partition of 56 genes into modules that maximized modularity $M_G$. This maximized modularity was used as the final modularity value for data analysis.

To calculate modularity of HCC samples grouped by metabolism phenotypes, tumor stages, metastatic potential, or recurrence status the RNA-seq data of each of the 56 AMPK and HIF-1 downstream genes the gene expression data were transformed by $log_2$ and normalization, i.e.

$$x \rightarrow \frac{log_2(x+1) - \overline{log_2(x+1)}}{\sigma(log_2(x+1))},$$

where $x$ represents the expression of each gene, $\overline{log_2(x+1)}$ is the mean of the $log_2$ transformed values across all patients' expression of this gene, and $\sigma(log_2(x+1))$ is the standard deviation of the $log_2$ transformed values.

The metabolic gene network for each group was defined by setting the 56 genes as the nodes and the Pearson correlation coefficient between genes as the link weights. The resulting network was represented by a 56*56 correlation matrix $C$. Since the above definition of modularity is for an unsigned graph, and since we regard negative correlations as weak links between genes, the whole matrix was shifted as $C' = (C+1)/2$. We used the Newman algorithm to calculate the modularity of this matrix $C'$.

To compare modularity between different patient groups, e.g. glycolysis versus OXPHOS, the bootstrapping method was used. This method takes the observed individual gene expression values as the most representative measure of the underlying distribution of expression values. That is, the distribution of expression values is taken as a sum over $\delta$ functions at the observed values. Predictions are computed from samples taken from this estimated distribution. For example, for the glycolysis group of 75 patients, the gene expression correlation matrix was calculated by randomly taking expression values from the 75 patients with replacement. The modularity of this correlation matrix was computed as described above. This sampling process was repeated 250 times to obtain 250 modularity values for the glycolysis group.

Mean modularity and standard error were then obtained. This same procedure was used to compute modularity for each of the other groups.

### 3. Calculation of Individual Modularity

Typically, for each patient there is one expression value for each gene, and no correlation between genes based upon only a single patient's data can be computed. We propose, therefore, to define the link between gene $i$ and gene $j$ of patient $\alpha$ as

$$l_{i,j}^{\alpha} = exp(-|X_{\alpha,i} - X_{\alpha,j}|/\sigma) \qquad (eq. 2)$$

where $X_{\alpha,i}$ is the expression of gene $i$ of patient $\alpha$, and $\sigma$ is the standard deviation of $|X_{\alpha,i} - X_{\alpha,j}|$ averaged across all pairs of genes and all patients, with $\sigma = 57\ 887$ in our case. This definition considers the link between gene $i$ and gene $j$ weak if the distance between them, i.e. $|X_{\alpha,i} - X_{\alpha,j}|$, is large. The scaling by $\sigma$ ensures that $|X_{\alpha,i} - X_{\alpha,j}|/\sigma$ remains within a reasonable order of magnitude.

Unlike the group modularity case, having only 56 expression values for each individual means the noise in the data has a greater impact on the calculated modularity values. Thus, a better way of filtering noise is needed. A standard approach is to reconstruct the data based only on cleaned leading eigenvectors. We utilized the iterative thresholding sparse PCA (ITSPCA) algorithm for this purpose [35] . The algorithm starts by keeping only the top eigenvectors. To separate signal and noise, such that signal is defined as above a threshold, a wavelet transformation is used (see Supplementary Materials and Methods section 2, Supplementary Fig. S1 and Supplementary Table S1). Data that were dense in real space became sparse in wavelet space, and a cutoff was then applied in wavelet space to eliminate the noise. The standard wavelet transformation algorithm requires that the number of entries be a power of two. Zero-padding was applied to the input data matrix $X$ so that it became a 371*64 matrix, with the last 8 columns containing only zeros. The ITSPCA algorithm output the cleaned version of leading eigenvectors $P_n$ (56*$n$ matrix), which can be used to reconstruct the raw data as

$$X' = X P_n P_n^T \qquad (eq.3)$$

where $X$ is the original raw data matrix, and $X'$ is the reconstructed matrix that has the same dimension of $X$. The cleaned data $X'$ should contain mostly signal and much less noise than $X$, and therefore $X'$ was used in calculation of links ($eq.2$). Note that, unlike the group modularity calculation, $X$ is based on the raw data without taking a logarithm. This is because we believe that noise had already been filtered out by ITSPCA, and taking the logarithm would only weaken the signal. See Supplementary Table S2 for chosen input parameters of the ITSPCA algorithm.

After determining $X'$, we computed the individual gene network linkage based on ($eq.2$). We then applied the binarization step where the top 5.6% edges (178 edges) were set to 1 and the rest set to zero. According to our previous work [30], this binarization step increases the signal-to-noise ratio without discarding important information. The Newman algorithm was used to compute modularity for each patient, $M_i$. We calculated the mean modularity of each patient group and the corresponding standard deviation of the mean to produce similar bar plots as in the group-level case.

**4. Definition of probability of surviving longer than 24 months based on individual modularity**

$$p_{survival,24}(M_i) = \frac{N_{survived,24} f_{survived,24}(M_i)}{N_{survived,24} f_{survived,24}(M_i) + N_{deceased,24} f_{deceased,24}(M_i)} \qquad (eq.4)$$

where $N_{survived,24}$ and $N_{deceased,24}$ are the numbers of patients that lived longer than 24 months and deceased within 24 months, respectively. Here $f_{survived,24}$ and $f_{deceased,24}$ are the probability density functions of the modularity distribution of survived and deceased group, respectively. Given modularity $M_i$, we calculated $p_{survival,24}$ and thus obtained the probability curve of surviving more than 24 months.

**5. Definition of probability of no recurrence in 12 months based on individual modularity**

$$p_{no\ recurrence,12}(M_i)$$
$$= \frac{N_{no\ recurrence,12} f_{no\ recurrence,12}(M_i)}{N_{no\ recurrence,12} f_{no\ recurrence,12}(M_i) + N_{recurrence,12} f_{recurrence,12}(M_i)} \quad (eq.5)$$

where $N_{no\ recurrence,12}$ and $N_{recurrence,12}$ are the numbers of patients that remained disease free for more than 12 months and those that recurred within 12 months, respectively. Here $f_{no\ recurrence,12}$ and $f_{recurrence,12}$ are the probability density functions of the modularity distribution of disease-free and recurred group, respectively. Given modularity $M_i$, we calculated $p_{no\ recurrence,12}$ and thus obtained the probability curve of no recurrence within 12 months.

**Conflict of Interest**

The authors declare no conflicts of interest.

**Funding**

Fengdan Ye, Dongya Jia, Herbert Levine and Michael Deem are supported by the National Science Foundation (NSF) grant PHY-1427654. Dongya Jia and Herbert Levine are additionally supported by the NSF grants DMS-1361411 and PHY-1605817. Herbert Levine was also supported by the Cancer Prevention and Research Institute of Texas (CPRIT) grants R1111. Mingyang Lu is partially supported by the National Cancer Institute of the National Institutes of Health grant P30CA034196.

# Figure Legends

**Figure 1.** Modular structure of the metabolic gene network. **A,** Evaluation of the AMPK and HIF-1 activities in HCC patients' samples (n = 371, r=-0.59, p<0.0001). Each point represents the AMPK and HIF-1 activities of one sample. **B,** Correlation matrix of the 33 AMPK downstream genes and 23 HIF-1 downstream genes. **C,** Rearranged correlation matrix calculated from the complete dataset of 371 HCC patients by the Newman algorithm. The Newman algorithm obtained a partition into two modules. Modules are labeled by black dashed lines.

**Figure 2.** Modularity and metabolism phenotypes. **A,** The 371 patients' samples are clustered into three metabolism phenotypes - OXPHOS (blue), hybrid (magenta) and glycolysis (red); **B,** Group modularity of three metabolism phenotypes; **C,** Kaplan-Meier (KM) overall survival curves.

**Figure 3.** Modularity and tumor stages. **A,** Bar plot of stage I and stage II-IV groups' modularity values. **B,** Kaplan-Meier (KM) overall survival curves.

**Figure 4.** Modularity and metastatic potential. Left panel: Group modularity of HCC samples with low and high metastatic potential evaluated by eight-gene signature (**A**) and SPP1 (**B**), respectively. Right panel: Kaplan-Meier (KM) overall survival curves of low and high metastatic potential groups on eight-gene signature (**A**) and SPP1 (**B**), respectively.

**Figure 5.** Modularity and tumor recurrence. **A,** Modularity (left panels) and Kaplan-Meier (KM) overall survival curves (right panels) of patients that were stratified into recurrence and no recurrence within 12, 24 and 36 months groups. **B,** Non-monotonic change of modularity with tumor recurrence time among patients that recurred. Group 1 is the longest from recurrence, and group 5 is the nearest. **C,** Kaplan-Meier (KM) overall survival curves of each group.

**Figure 6.** Individual modularity. **A-C,** Individual modularity results show the same trend of modularity with metabolism types, stages and metastatic potential. Pearson correlation between individual modularity and eight-gene metastatic potential r=0.46, p<1e-20. **D,** Left: probability of survival longer than 24 months derived from data. Middle: Gaussian distribution of modularity values for the two groups. Right: same probability based on Gaussian model. **E,** Left: probability of no recurrence in 12 months derived from data. Middle: Gaussian distribution of modularity values for the two groups. Right: same probability based on Gaussian model.

# Figures

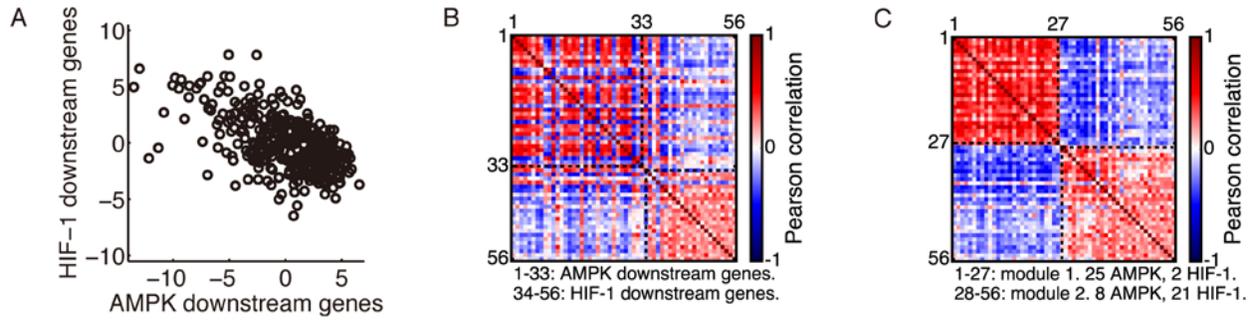

Figure 1

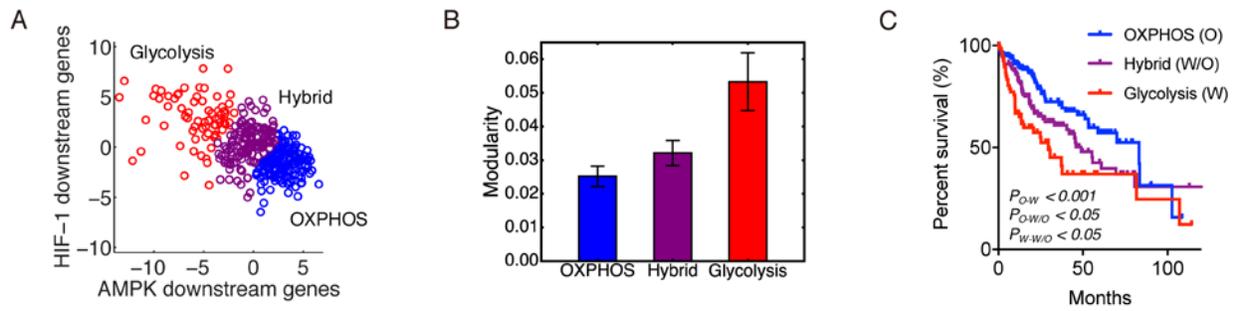

Figure 2

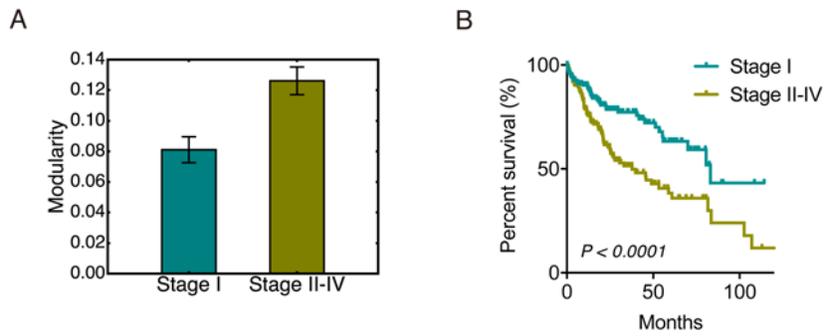

Figure 3

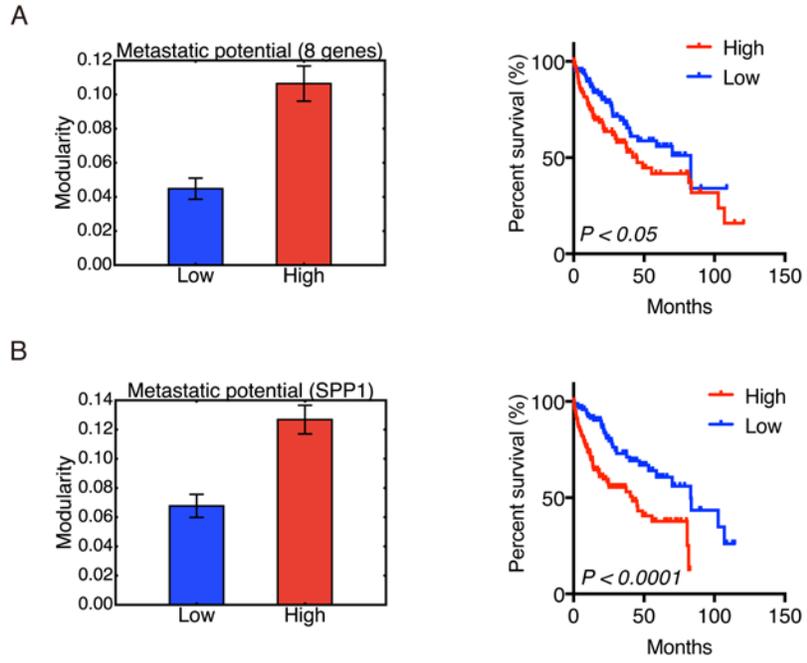

Figure 4

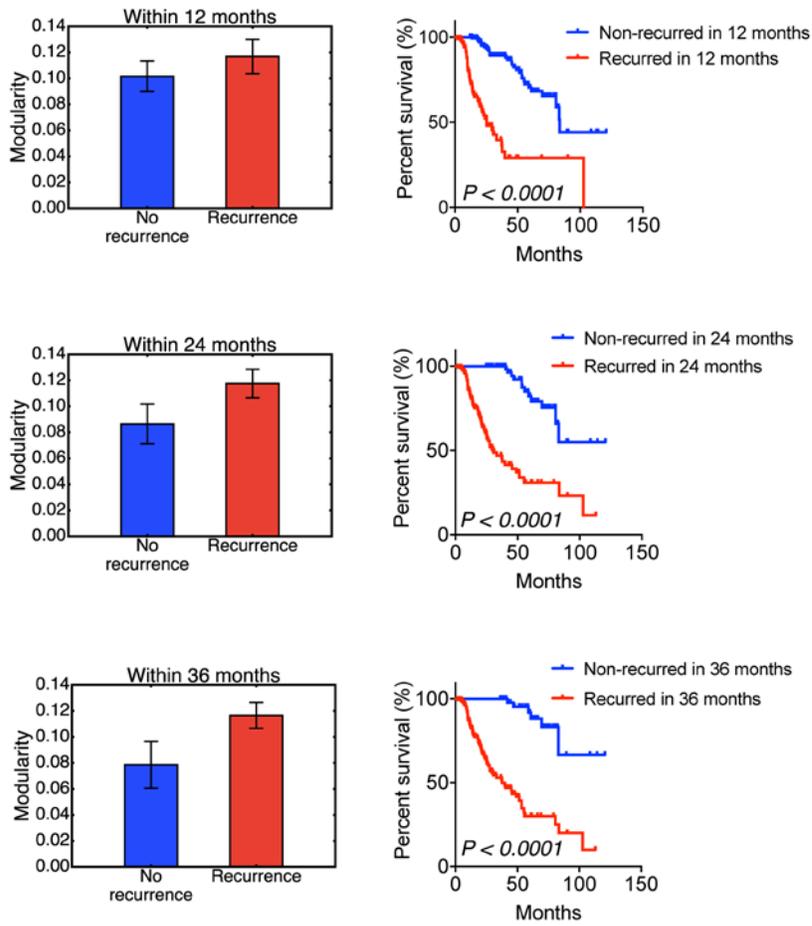
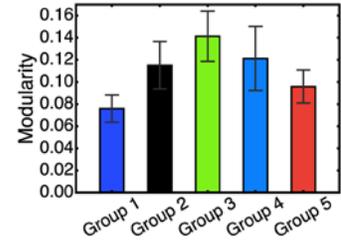
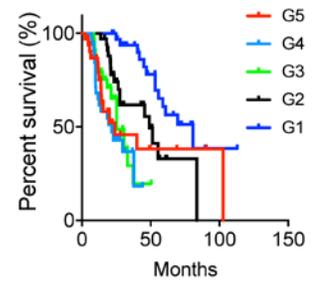

Figure 5

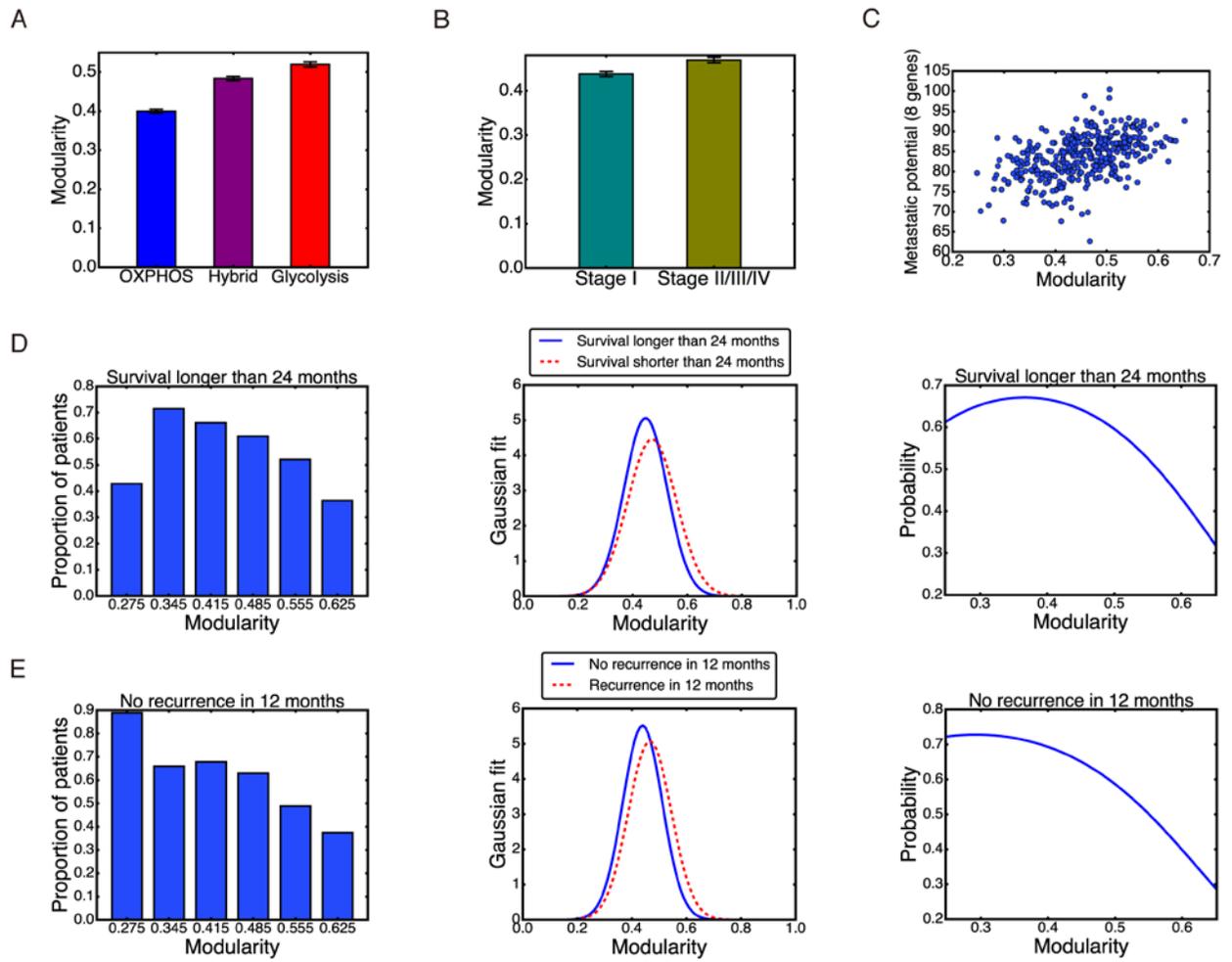

Figure 6

**Modularity of the metabolic gene network as a prognostic biomarker for hepatocellular carcinoma**

Fengdan Ye, Dongya Jia, Mingyang Lu, Herbert Levine, Michael W Deem

**Supplementary Tables**

**Table S1.** DWT frequency domain of an example of signal $x[n]$ with a size of 64 under a 3-level decomposition. Frequency range of the signal is 0 to $f$.

| Level | Frequency range | Size of sample |
|---|---|---|
| 1 | $f/2$ to $f$ | 32 |
| 2 | $f/4$ to $f/2$ | 16 |
| 3 | $f/8$ to $f/4$ | 8 |
| 3 | 0 to $f/8$ | 8 |

**Table S2.** Input parameters of the ITSPCA algorithm.

| Parameter | Value |
|---|---|
| Number of significant leading eigenvectors $n$ | 10 |
| Wavelet basis to be used | Symmlet |
| Coarsest level in wavelet transform $L$ | 4 |
| Parameter describing the support length and vanishing moments of the selected wavelet basis, *par* | 8 |
| Adjustable constant in the diagonal thresholding step, $\alpha$ | 3 |
| Adjustable constant in the iterative thresholding steps, $\beta$, which is directly related to level sparsity in cleaned eigenvectors. | 1.5 |
| Thresholding rule, *itthres*. | hard |